\documentclass[10pt]{iopart} 
\usepackage[utf8]{inputenc}
\usepackage[english]{babel}
\usepackage{footmisc}
\expandafter\let\csname equation*\endcsname\relax
\expandafter\let\csname endequation*\endcsname\relax
\usepackage{amsmath}
\usepackage{iopams}
\usepackage{graphicx}
\usepackage{upgreek} 
\usepackage{url}
\usepackage{float}
\usepackage{verbatim} 
\usepackage{array}
\usepackage{caption}
\usepackage{subcaption}

\usepackage{ulem} 

\newcommand{\hm}[1]{{#1}}
\usepackage{todonotes}

\newcommand\Vappl{V$_{\rm appl}$}

\newcommand{\tinter}{\Delta \textrm{t}_\textrm{inter}}

\newcommand\micros{$\upmu$s}

\begin{document}
\title[RPD, version of \today]{Double-pulse streamer simulations for varying interpulse times in air }
\author{H. Malla$^1$, A. Martinez$^{1, 2}$, U. Ebert$^{1,2}$ and J. Teunissen$^1$}
\address{$^1$ Centrum Wiskunde \& Informatica (CWI), Amsterdam, The Netherlands}
\address{$^2$ Department of Applied Physics, Eindhoven University of Technology, PO Box 513, 5600 MB Eindhoven, The Netherlands}
\eads{\mailto{jannis.teunissen@cwi.nl}}

\begin{abstract}
  In this paper, we study how streamer discharges are influenced by a previous voltage pulse \hm{using an axisymmetric fluid model}.
  \hm{We simulate double-pulse positive streamers in N$_2$-O$_2$ mixtures containing 20\% and 10\% O$_2$ at 1 bar.}
  By varying the time between the pulses between 5\,ns and 10 $\mu$s, we observe three regimes during the second pulse: streamer continuation, inhibited growth and streamer repetition.
  In the streamer continuation regime, a new streamer emerges from the tip of the previous one.
  In the inhibited regime, the previous channel is partially re-ionized, but there is considerably less field enhancement and almost no light emission.
  Finally, for the longest interpulse times, a new streamer forms that is similar to the first one.
  The remaining electron densities at which we observe streamer continuation agree with earlier experimental work.
  We introduce an estimate which relates streamer continuation to the dielectric relaxation time, the background field and the pulse duration.
  Furthermore, we show that for interpulse times above 100 ns several electron detachment reactions significantly slow down the decay of the electron density.


\end{abstract}


\ioptwocol
\maketitle
\section{Introduction}

Streamers are transient, filamentary gas discharges~\cite{vitello1994simulation,nijdam_physics_2020}.
Due to the strong electric field enhancement at their tips they can propagate in background electric fields below the breakdown threshold.
In nature they occur as streamer coronas ahead of lightning leaders, and as tens of kilometers tall sprite discharges in the thin atmosphere high above thunderstorms.
Streamer discharges can be used to generate various chemical species and they are used in applications such as plasma medicine~\cite{verloy_cold_2020, graves_low_2014, laroussi_2014_medicine}, agriculture~\cite{ranieri_plasma_2021}, industrial surface treatments~\cite{bardos2010cold}, and combustion~\cite{starikovskaia2006plasma}.

Streamers are commonly produced using a repetitively pulsed voltage source \cite{nijdam_investigation_2014,li2018positive,mirpour_distribution_2020}.
The repetition rate determines how strongly discharges are affected by previous pulses, through left-over ionized and neutral species as well as gas heating.
Such effects from previous pulses can cause transitions between corona, glow and spark discharge regimes~\cite{pai2010transitions}.
In~\cite{tholin2013simulation}, simulations were performed to study the evolution of electrons, positve and negative ions during the interpulse.

The effect of pulse repetition frequencies on positive streamers in molecular nitrogen-oxygen mixtures was studied systematically in double pulse experiments in~\cite{nijdam_investigation_2014, li2018positive}.
In~\cite{nijdam_investigation_2014} different formation and propagation behavior was observed for the second-pulse streamer depending on the time between the two voltage pulses.
If this time was sufficiently short, the first-pulse streamers continued their growth during the second pulse.
The maximum time between two voltage pulses for which this happens was called the continuation time.
The authors were able to explain how continuation times vary with the \hm{O$_2$} percentage by zero-dimensional plasma-chemical modeling.

The effect of previous pulses has also been analyzed experimentally or using global kinetics models in the context of plasma jets in~\cite{lu2018guided, chang2016effect}.
They focus on the minimum electron density needed to observe the repeatable behaviour of a plasma jet.
In a recent paper~\cite{babaeva2022evolution}, three consecutive negative voltage pulses were simulated for an atmosperic plasma jet to investigate the role of left-over charged species after the first pulse.
\hm{Repetitive-pulsed DBD experiments were performed in~\cite{hoft2014bidirectional,hoft2014breakdown}, and different breakdown regimes were identified depending on the pulse-off time similar to~\cite{nijdam_investigation_2014}.}
Furthermore, streamer memory effects and discharge mode transitions due to repetitive pulses have been studied in a series of papers by Zhao~\textit{et al}, see e.g.~\cite{Zhao_2020,Zhao_2023}, and reviewed in~\cite{Zhao_2020a}.

The remaining electron density from a previous pulse can influence where a discharge will grow during the next pulse.
Such an effect was observed~\cite{nijdam_investigation_2014}, and it was studied in another context with laser-generated ionization in~\cite{nijdam_streamer_2014,Nijdam_2016a}.
The contribution of electrons detached from various negative ions to the seed electrons of subsequent pulses was studied in several papers~\cite{pancheshnyi2005role,naidis2011modelling,babaeva2022evolution,chang2016effect}.

In this paper, we address an open question posed in~\cite{nijdam_investigation_2014}: which physical mechanism determines the electron density required for streamer continuation?
Furthermore, we study how the second-pulse discharge changes when the time between the voltage pulses is increased beyond the continuation time.
To this end, we perform simulations of double-pulse positive streamers in \hm{two N$_2$-O$_2$ mixtures with 20\% and 10\% O$_2$} at standard temperature and pressure, using a 2D axisymmetric fluid model.
We use a detailed plasma chemistry including neutral and excited species, which allows us to investigate the relative contribution of various detachment reactions after the first pulse.
\section{Model}
\label{sec:model}

We study streamer discharges at 1\,bar and 300 K in two nitrogen-oxygen mixtures containing 20\% and 10\% oxygen in an average background field of 15\,kV/cm.
Axisymmetric simulations are performed with a standard drift-diffusion fluid model with the local field approximation, using the \texttt{afivo-streamer} code \cite{teunissen_simulating_2017}.
For a recent comparison of this model against experiments and against particle simulations see \cite{li_comparing_2021} and \cite{Wang_2022}.
We briefly summarize the main equations below, for further details see e.g.~\cite{teunissen_simulating_2017,Bagheri_2018a}.

\subsection{Equations}
\label{sec:equations}

The electron density $n_e$ evolves in time as
\begin{equation}\label{eq:dd_electrons}
  \partial_t n_e = \nabla \cdot (\mu_e \mathbf{E} n_e + D_e \nabla n_e) + S_e + S_{\mathrm{ph}},
\end{equation}
where $\mu_e$ is the electron mobility coefficient, $D_e$ the diffusion coefficient, and $\mathbf{E}$ the electric field.
\hm{The electron transport coefficients are assumed to depend on the local electric field (local field approximation) and were computed using BOLSIG-~\cite{hagelaar_solving_2005} using Phelps' cross-section data~\cite{Phelps_database}.}
$S_e$ is a source (and sink) term due to reactions involving electrons, for example ionization or attachment, see section \ref{sec:reaction_set}.
$S_{\mathrm{ph}}$ is a photo-ionization source term, which is here implemented according to Zhelezniak's model~\cite{zheleznyak1982photoi_english} as discussed in \cite{Bagheri_2018a}.
Ion \hm{and neutral} densities $n_i$ (numbered by $i=1,\ldots,n$) evolve in time as
\begin{equation}\label{eq:dd_ions}
  \hm{\partial_t n_i + \nabla \cdot (q_i \mu_i \mathbf{E} n_i) = S_i.}
\end{equation}
Here $S_i$ is a source/sink term due to reactions, $\mu_i$ is the mobility and $q_i$ accounts for the species' charge \hm{(0 for neutrals, $-1$ for negative ions and $+1$ for positive ions)}.

The electric field $\mathbf{E}$ is calculated in the electrostatic approximation as $\mathbf{E} = -\nabla \phi$.
Here $\phi$ is the electrostatic potential, which is obtained by solving the Poisson equation~\cite{Teunissen_2018_afivo,Teunissen_2023}
\begin{equation}
  \nabla^2 \phi = -{\rho}/{\epsilon_0},
\end{equation}
where $\rho$ is the space charge density and $\epsilon_0$ is the vacuum permittivity.

\subsection{Geometry}
\label{sec:geom}

\begin{figure}
    \centering
    \includegraphics[width=8cm]{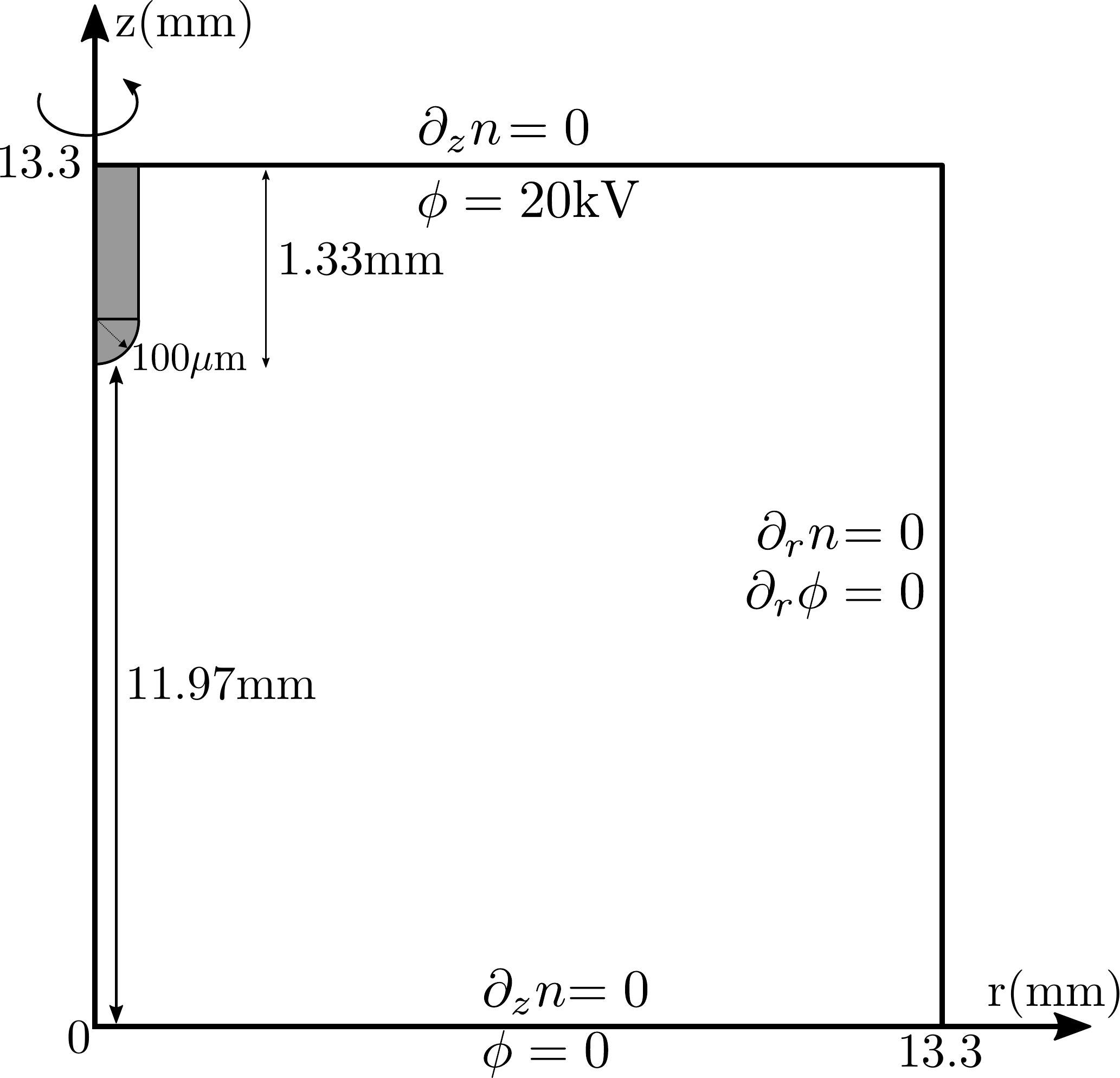}
    \caption{The axi-symmetric plate-plate geometry with a needle electrode.
      Boundary conditions for the species densities $n$ and the electrostatic potential $\phi$ for each of the domain boundaries are given.}
    \label{fig:geometry}
  \end{figure}

We use a plate-plate geometry with a needle electrode which is shown in figure \ref{fig:geometry}.
For species densities, homogeneous Neumann boundary conditions are used on all domain boundaries.
For the electric potential, a homogeneous Neumann boundary is used on the outer radial boundary.
The bottom plate is grounded, while a high voltage is on the upper plate and the needle electrode.
The applied voltage is 20\,kV, so that the average field between the plates is 15\,kV/cm, which is about half the breakdown field at 1\,bar.

\subsection{Voltage waveform and initial conditions}
\label{sec:voltage}

\begin{figure}
    \centering
    \includegraphics[width=8cm]{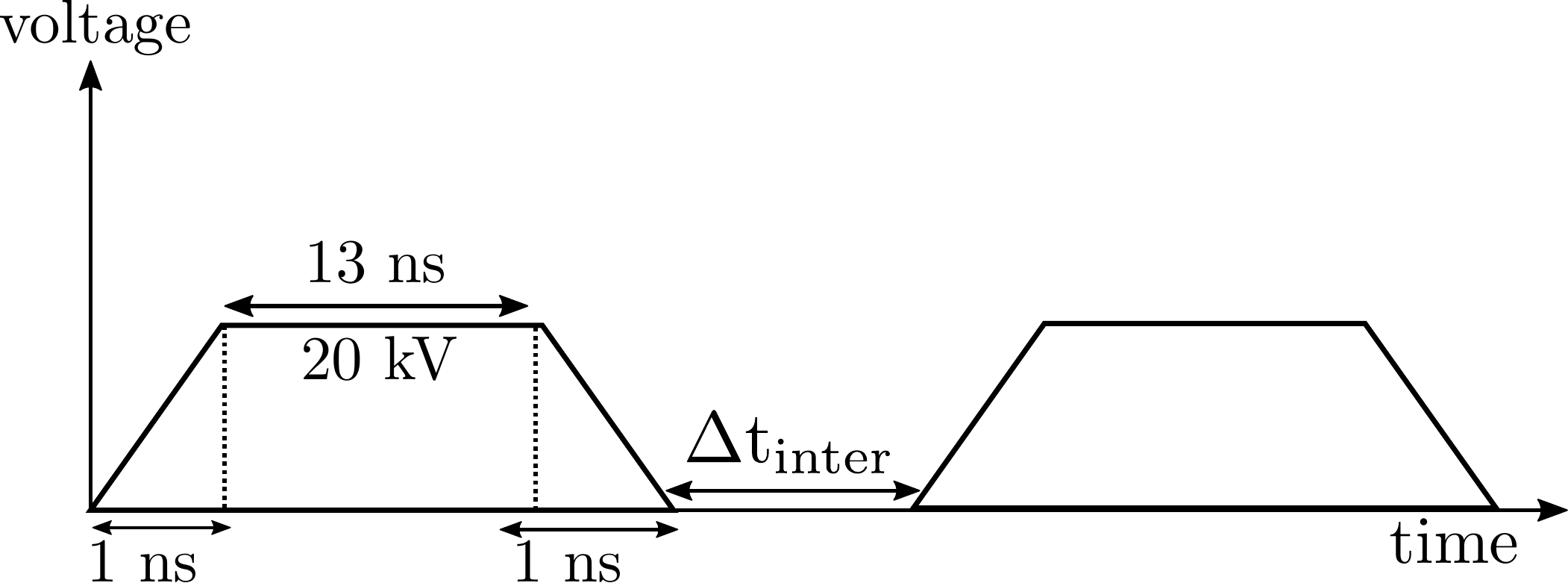}
    \caption{The applied voltage waveform with two identical pulses.
    The time between the pulses $\Delta t_\textrm{inter}$ is varied from 5~ns to 10~$\mu$s.}
    \label{fig:voltage_shape}
\end{figure}

The applied voltage waveform is shown in figure \ref{fig:voltage_shape}, consisting of two identical pulses separated by a time $\tinter$ that is varied from 5\,ns to 10\,\micros{}.
The peak voltage \Vappl{} is 20\,kV, which is applied during 13\,ns, and the rise and fall times of the voltage are 1\,ns.
As an initial condition, we use a neutral (electrons and O$_2^+$) background ionization density of $1.8~\times~10^{9}$\,m$^{-3}$.
\hm{This background ionization provides the first electrons for streamer inception near the electrode tip, where there is significant electric field enhancement.
Inception occurs approximately 1\,ns after the voltage is turned on.}

\subsection{Reaction Set}
\label{sec:reaction_set}

We use the reaction set given in~\cite{baohong_chemistry}, containing 263 reactions, which was primarily compiled from the reactions given in~\cite{kossyi_kinetic_1992, ono2020}.
This set includes reactions between electrons, neutrals, ions, and excited species.
A list of the 56 considered species is given in table \ref{tab:species}.
Rate constants for reactions involving electron collisions were computed with BOLSIG$-$~\cite{hagelaar_solving_2005} using the cross-sections from \cite{Phelps_database}.
\hm{We remark that the reaction set was designed for dry air (20\% O$_2$, 80\% N$_2$), and that using the same reactions for 10\% O$_2$ is an approximation as some rate coefficients could depend on the O$_2$ concentration.}

We consider the motion of seven major ion species: N$_2^+$, O$_2^+$, N$_4^+$, O$_4^+$, O$_2^-$, O$^-$, and  O$_3^-$; \hm{other ions are assumed to be immobile.}
\hm{For simplicity, we assume that the mobile ions all have a constant mobility $2.2 \times 10^{-4} \text{m}^2 \text{V}/ \text{s}$~\cite{tochikubo2002numerical}, because ion motion played no major role for the results reported in this paper.}

In section~\ref{sec:diffchemset}, we use the reaction set used in~\cite{li2022computational} which is a subset of the above described reaction set to demonstrate the effect of using a simplified reaction set on the interpulse plasma evolution.

\begin{table}
    \caption{\hm{List of species used in the simulations. We use the reaction set described in~\cite{baohong_chemistry}.}}
    \centering
    \begin{tabular}{c | p{0.92\linewidth}}
        Neutral & N$_2$, O$_2$, NO, NO$_2$, NO$_3$, N$_2$O, N$_2$O$_3$, N$_2$O$_5$, N$_2$O$_4$, O$_3$, N($^4$S), N($^2$D), N($^2$P), O($^1$D), O($^1$S), O($^3$P)   \\
        Positive & N$^+$, N$_2^+$, N$_3^+$, N$_4^+$, N$_2$O$_2^+$, N$_2$O$^+$, NO$_2^+$, NO$^+$, O$_2^+$, O$_4^+$, O$^+$ \\
        Negative & e$^-$, N$_2$O$^-$, NO$_2^-$, NO$_3^-$, NO$^-$, O$_2^-$, O$_3^-$, O$_4^-$, O$^-$ \\
        Excited & N$_2(\text{rot})$, O$_2(\text{rot})$, N$_2$(A), N$_2$(B), N$_2$(a), N$_2$(C), N$_2$(E), O$_2$(a), O$_2$(b), O$_2$(A), N$_2$(v=1-8), O$_2$(v=1-4)
    \end{tabular}
    \label{tab:species}
  \end{table}

\section{Results}
\label{sec:results}

\subsection{Effect of interpulse time on streamer continuation}
\label{sec:interpulse_continuation}
\begin{figure*}
  \centering
  \includegraphics[width=16cm]{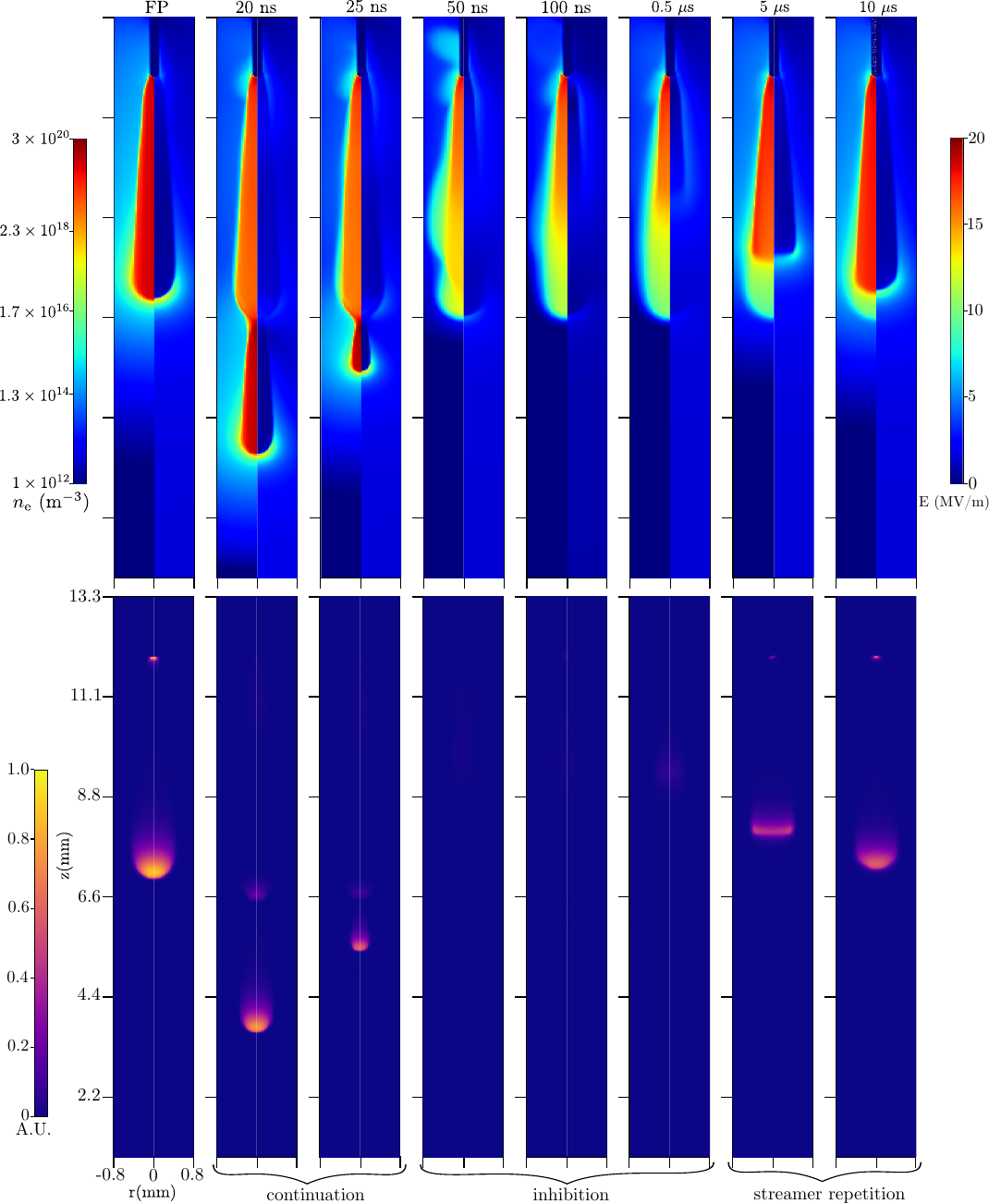}
  \caption{Plots of the first pulse (FP) streamer (left most column) and of the second streamer for interpulse times of 20 ns to 10 $\mu$s (as indicated above each column), for $20\%$\,O$_2$.
    Each column shows electron density (top, left half) and electric field (top, right half) and instantaneous light emission (bottom), all at the end of the pulse but before the fall time.
    \hm{Light emission was computed using a forward Abel transform.}
    The three observed regimes, streamer continuation, inhibited growth and streamer repetition are indicated.
  }
  \label{fig:20O2_ne_le_efld}
\end{figure*}
\begin{figure*}
  \centering
  \includegraphics[width=16cm]{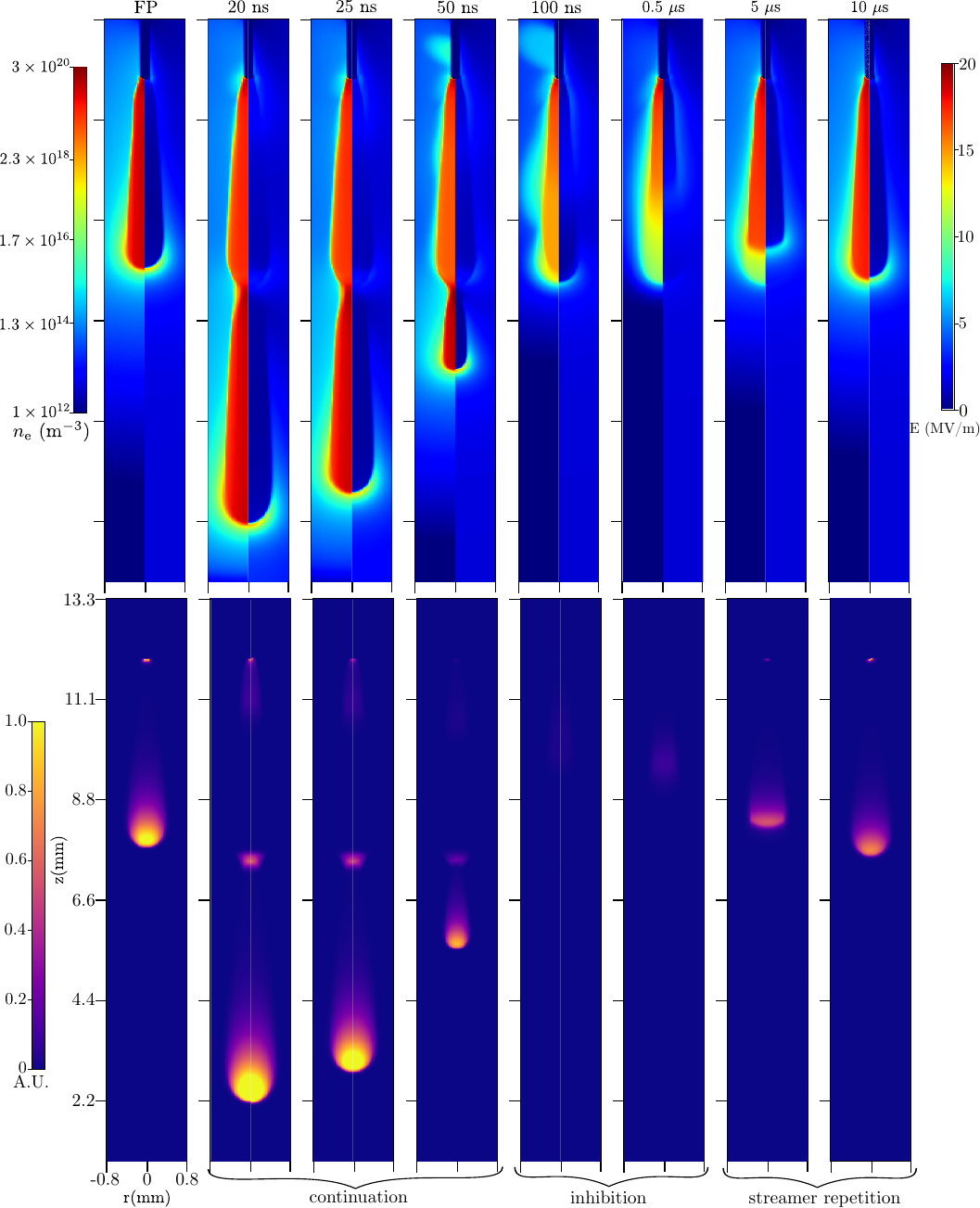}
  \caption{The same plots as in figure~\ref{fig:20O2_ne_le_efld}, but now for 10\% O$_2$.
  }
  \label{fig:10O2_ne_le_elfd}
\end{figure*}
We vary the interpulse time $\tinter$ in two gas mixtures, consisting of N$_2$ with 20\% O$_2$ or 10\% O$_2$, to study the properties of the streamer generated by the second voltage pulse.
Figures~\ref{fig:20O2_ne_le_efld} and~\ref{fig:10O2_ne_le_elfd} show streamers at the end of the second pulse (but before the voltage fall time) for varying $\tinter$, for the cases of 20\% O$_2$ and 10\% O$_2$, respectively.
In these figures, the instantaneous light emission is approximated by the \hm{Abel-transformed} N$_2(\mathrm{C}^3\Pi_u)$ density, since most of the emission comes from the second positive system~\cite{pancheshnyi_development_2005}.
\hm{The forward Abel transform was performed using the Hansen-Law method~\cite{hansen1985recursive}.}
Axial electric field profiles corresponding to figure~\ref{fig:20O2_ne_le_efld} are shown in figure \ref{fig:20O2_axialEfld}.

Depending on $\tinter$, we observe three regimes: continuation, inhibited growth and streamer repetition, which are discussed below.
In figure~\ref{fig:20O2_secondPulseevolution}, we show the streamer evolution during the second pulse at equally spaced time intervals for $\tinter$ = 25 ns and 50 ns.

\subsubsection{Continuation regime.}
For short interpulse times the channel created by the first pulse still has a relatively high conductivity.
The existing channel therefore becomes electrically screened during the second pulse, as shown in figure~\ref{fig:20O2_secondPulseevolution}a, leading to the emergence of a new streamer from its tip.
For larger $\tinter$ the conductivity of the existing channel is lower, which results in a longer electric screening time.
This results in a longer inception delay for the second streamer, which is therefore shorter.
The maximum time between the pulses for which continuation occurs is here called the streamer continuation time.
In section \ref{sec:criterion-streamer-continuation} a criterion for streamer continuation is discussed.

\subsubsection{Inhibited regime.}

\begin{figure*}
  \centering
  \includegraphics[width=16cm]{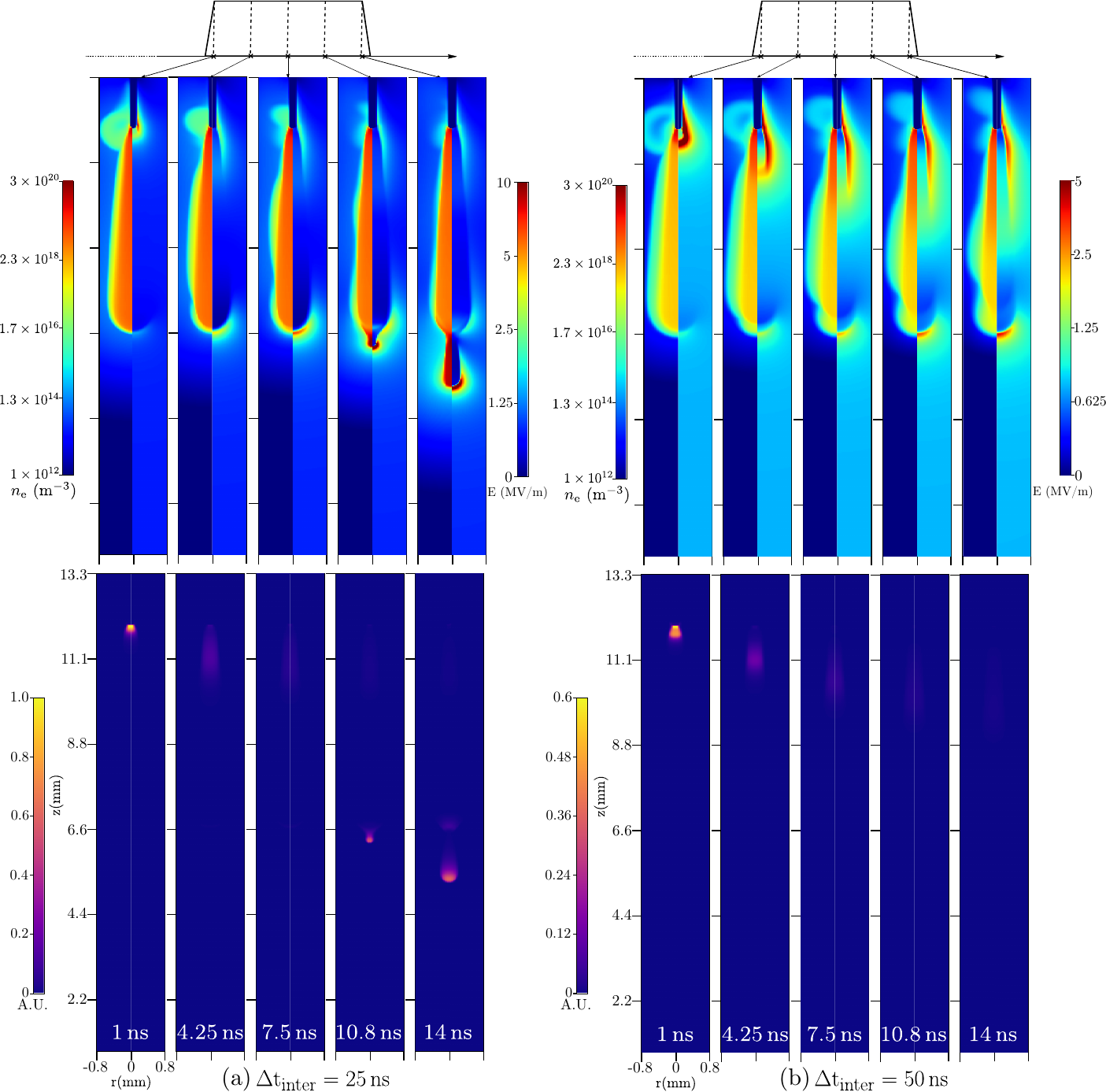}
  \caption{Evolution during the second voltage pulse for interpulse times of 25 and 50 ns, for $20\%$\,O$_2$, at equally spaced intervals.
    The time offsets with regard to the rise of the second voltage pulse are indicated at the bottom and visually at the top.
    \hm{On top, the electron density and the electric field are shown (left and right half of each image), and on the bottom the instantaneous light emission is shown.}}
  \label{fig:20O2_secondPulseevolution}
\end{figure*}

For longer $\tinter$ the first-pulse streamer channel has lost so much of its conductivity that it does not become fully screened during the second pulse (see figure~\ref{fig:20O2_secondPulseevolution}b), and there is no streamer continuation. 
However, the electron density in the old channel is still high enough to inhibit a new streamer from forming, and a weak ionization wave passes through the channel instead.
Light emission plots in figure~\ref{fig:20O2_secondPulseevolution} show that there is almost no light emission during the second pulse.
In~\cite{nijdam_investigation_2014}, such an inhibited regime was also observed.
Other studies have also found that a relatively high background electron density reduces the field enhancement of a streamer, leading to slower discharge growth and a lower degree of ionization, see e.g.~\cite{Nijdam_2016a,li_comparing_2021}.

\subsubsection{Streamer repetition regime.}

\begin{figure*}
  \centering
  \includegraphics[width=12cm]{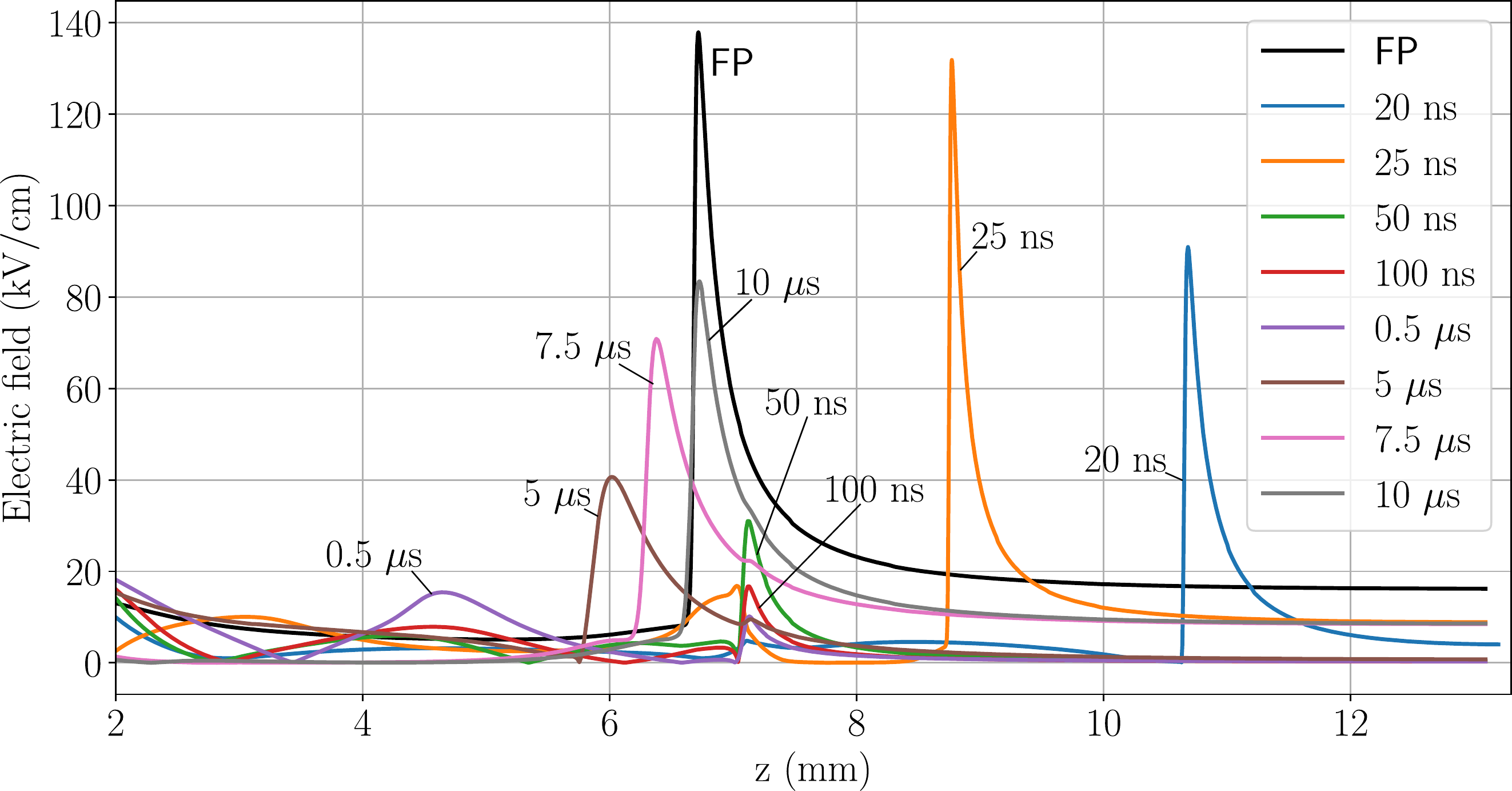}
  \caption{Electric field profile at the end of the second pulse (but before the voltage falls) for 20\% O$_2$, at varying interpulse times. Electric field profile for the first-pulse streamer is plotted as a black curve for reference.}
  \label{fig:20O2_axialEfld}
\end{figure*}

For even longer $\tinter$, the re-ionization of the channel becomes more streamer-like, as can be seen in figure~\ref{fig:20O2_axialEfld}: the electric field is enhanced at the tip of the re-ionized region and it is screened behind the tip.
Note that the electric field for this second-pulse streamer is still weaker than for the first pulse, and that its head shape is somewhat deformed.
Eventually, the electron density of the previous channel will become so low that it hardly affects the evolution during the second pulse.
We approach this behavior for $\tinter = 10\,\mu \mathrm{s}$, where the second-pulse streamer therefore closely resembles the first-pulse streamer.

\hm{One might expect the effect of gas heating to influence the second-pulse streamer properties.
However, this is not the case beause the amount of energy deposited before the second pulse is so small that ambient gas heating is negligible, as discussed in~\ref{sec:gas_heating}.}



\subsection{Channel evolution between pulses}
\label{sec:first-pulse-streamer-interpulse}

\begin{figure}
  \centering
  \includegraphics[width=8cm]{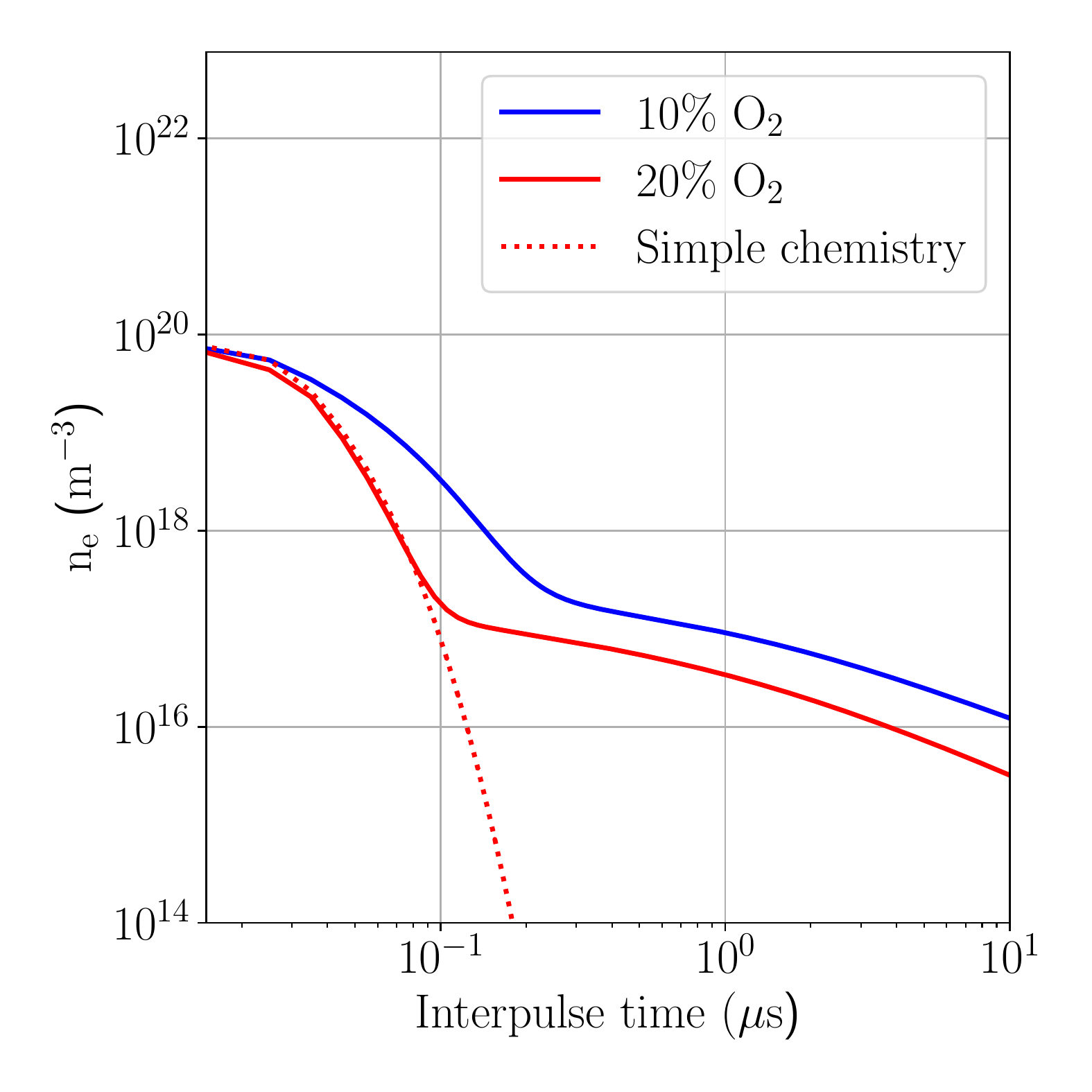}
  \caption{The temporal evolution of the electron density during the interpulse phase.
    The values were obtained at a point $1 \, \textrm{mm}$ below the electrode tip.
    The plot with ``Simple chemistry" will be elaborated further in section~\ref{sec:diffchemset}
  }
  \label{fig:interpulse_pvt}
\end{figure}

To better understand the evolution of second-pulse streamer, figure~\ref{fig:interpulse_pvt} shows the time evolution of the electron density at a point 1 mm below the electrode tip starting from the moment when the first pulse is off, i.e., from t\,=\,15 ns.
We observe two electron decay timescales.
For times up to 0.1$\, \mu$s, the electron density and hence the conductivity decay rapidly due to electron attachment and electron-recombination reactions.
For t$\, > \, 0.1 \, \mu$s, the electron decay rate reduces due to detachment reactions.
The main attachment and detachment reactions while the voltage is off are given in table \ref{tab:ereactions_interpulse}.

\hm{The concentration of O$_2$ in the ambient gas mixture affects the relative contribution of different attachment reactions.}
For example, during an interpulse time of 500\,ns, the percentage contribution to electron loss is as follows:
\begin{itemize}
  \item Three-body attachment to O$_2$ gas (R1+R2): 75\% (20\%\,O$_2$) and 55\% (10\%\,O$_2$).
  Note that the rates of R1 and R2 depend quadratically and linearly on the O$_2$ concentration.
  \item Dissociative recombination (R3): 24\% (20\%\,O$_2$) and 44\% (10\%\,O$_2$).
  This reaction results from the fast conversion of positive ions into O$_4^+$ inside the streamer channel~\cite{aleksandrov_ionization_1999}.
\end{itemize}
Other reactions are responsible for only about 1\% of the electron loss.
These relative contributions are not sensitive to the interpulse time, with a variation of less than 10\% for all the interpulse times.

\begin{table*}
  \centering
  \captionsetup{width=0.92\textwidth}
  \caption{Major reactions that consume and produce electrons when the voltage is turned off.
    \hm{The full reaction set as specified in~\cite{baohong_chemistry} and used throughout the paper contained 21 electron attachment and recombination reactions and 18 electron detachment reactions.}
    \hm{Reaction R4 from~\cite{pancheshnyi2013effective} was included to account for field-dependent electron detachment. Note the rate of this reaction should depend on the O$_2$ concentration~\cite{pancheshnyi2013effective}, but for simplicity we here use a single rate coefficient for both O$_2$ concentrations, as the rates are similar.}
  $T$ (K) and $T_e$ (K) are gas and electron temperatures, respectively.
  $T_e$ is computed as $T_e = 2\epsilon_\mathrm{e} / 3k_\mathrm{B}$ with the mean electron energy $\epsilon_\mathrm{e}$ obtained from BOLSIG+~\cite{Bolsig_application}. The ``simple chemistry" added in Fig.~\ref{fig:interpulse_pvt} also consists of reactions R1, R3, and R4.}
  \begin{tabular*}{0.92\textwidth}{c@{\extracolsep{\fill}}llc}
  \br
  No. & Reaction & Reaction rate coefficient($\mathrm{m^6\,s^{-1}}$ or $\mathrm{m^3\,s^{-1}}$) & Reference\\
  \mr
  \multicolumn{4}{l}{Major electron-loss reactions}\\
  R1 & 
  $\mathrm{e} + \mathrm{O}_2 + \mathrm{O}_2 \to \mathrm{O}_2^- + \mathrm{O}_2$
 & $k_{1}(E/N)$ & \cite{Phelps_database}\\
  R2 & 
  $\mathrm{e} + \mathrm{O}_2 + \mathrm{N}_2 \to \mathrm{O}_2^- + \mathrm{N}_2$
 & $k_{2}=10^{-43}$ & \cite{Phelps_database}\\
  R3 & $\mathrm e + \mathrm O_4^+ \to \mathrm O_2 + \mathrm O_2$ & $k_{3}=1.4\times10^{-12}(300/T_e)^{0.5}$ & \cite{kossyi_kinetic_1992} \\
  \mr
  \multicolumn{4}{l}{Electron-detachment reactions}\\
  \br
  R4 & $\mathrm O_2^- + \mathrm M \to \mathrm e + \mathrm O_2 + \mathrm M$ & $k_{4}=1.24\times10^{-17}\exp(-(\frac{179}{8.8+E/N})^2)$ & \cite{pancheshnyi2013effective}\\
  R5 & $\mathrm O_2^- + \mathrm O_2^*(b) \to \mathrm e + 2\mathrm O_2$ & $k_{5}=3.60\times10^{-16}$ & \cite{kossyi_kinetic_1992}\\
  R6 & $\mathrm O_2^- + \mathrm N_2^*(A) \to \mathrm e + \mathrm O_2 + \mathrm N_2$ & $k_{6}=2.10\times10^{-15}$ & \cite{kossyi_kinetic_1992}\\
  R7 & $\mathrm O_2^- + \mathrm O(^3\mathrm P) \to \mathrm e + \mathrm O_3$ & $k_{7}=1.50\times10^{-16}$ & \cite{kossyi_kinetic_1992}\\
  R8 & $\mathrm O_2^- + \mathrm N(^4\mathrm S) \to \mathrm e + \mathrm N \mathrm O_2$ & $k_{8}=5\times10^{-16}$ & \cite{kossyi_kinetic_1992}\\
  R9 & $\mathrm O_3^- + \mathrm O(^3\mathrm P) \to \mathrm e + 2\mathrm O_2$ & $k_{9}=3\times10^{-16}$ & \cite{kossyi_kinetic_1992}\\
  \label{tab:ereactions_interpulse}
  \end{tabular*}
\end{table*}

\section{Discussion}
\label{sec:discussion}

\subsection{Comparison with earlier experimental work}

In~\cite{nijdam_streamer_2014}, double pulse streamer experiments were performed at 133 mbar for varying interpulse times and for varying \hm{O$_2$} concentrations.
With 0D plasma-chemical modeling, it was estimated that the minimum remaining electron density $n_e^ \mathrm{min}$ for streamer continuation was about $5\times 10^{17}$m$^{-3}$, using a pulse duration $t_\mathrm{const}$ of about $200-300$\,ns with a rise/fall time of 15\,ns.
At 1 bar, these values scale to $n_e^ \mathrm{min} \sim 3\times 10^{19}$m$^{-3}$ and $t_\mathrm{const} \sim 27-30 \, \mathrm{ns}$, see~\cite{nijdam_physics_2020}.

For the simulations performed here, with a pulse duration $t_\mathrm{const} \sim 13 \, \mathrm{ns}$, we observe streamer continuation for $n_e^ \mathrm{min}$ in the range of $10^{19}  \, \mathrm{m}^{-3}$ up to $4\times 10^{19} \, \mathrm{m}^{-3}$.
Compared to the results of~\cite{nijdam_streamer_2014}, some deviations are to be expected, due the uncertainties in the 0D plasma-chemical modeling in~\cite{nijdam_streamer_2014}, due to corrections to the scaling with pressure, and due to the different electrode geometries and pulse shapes.
We therefore consider our results to be in relatively good agreement with~\cite{nijdam_streamer_2014}.

\hm{In~\cite{hoft2014breakdown}, experiments were performed in a 1 mm overvolted DBD gap containing N$_2$ with 0.1\% O$_2$ at atmospheric pressure, and a 10\,kV voltage was applied with a 10\,kHz repetition frequency.
  Different breakdown regimes were observed by varying the voltage-off time $t_\mathrm{off}$ between pulses.
  For $t_\mathrm{off} > 20 \, \mu\mathrm{s}$, a positive streamer propagated between the dielectric-covered electrodes, similar to the streamer repetition regime shown in figures~\ref{fig:20O2_ne_le_efld}--\ref{fig:10O2_ne_le_elfd}.
  For $20 \, \mu\mathrm{s} > t_\mathrm{off} > 4 \, \mu\mathrm{s}$, a slower and more diffuse positive streamer developed, which was attributed to a high residual electron density in the gap.
  Finally, for $t_\mathrm{off} < 0.5 \, \mu\mathrm{s}$ no streamer propagation occurred, and instead a re-ignition of the previous discharge's afterglow was observed, similar to the inhibition regime shown in figures~\ref{fig:20O2_ne_le_efld}--\ref{fig:10O2_ne_le_elfd}.
  The authors explain these regimes through 1D modeling, which correlates $t_\mathrm{off}$ with the background ionization density that is present at the start of the next pulse, similar to the present study.
  The fact that similar regimes were observed despite the rather different operating conditions shows that left-over ionization plays a generic role in repetitively pulsed discharges.
}

\subsection{Criterion for streamer continuation}
\label{sec:criterion-streamer-continuation}

If the channel eventually becomes electrically screened during the second pulse, as illustrated in the evolution in figure~\ref{fig:20O2_secondPulseevolution}, a high electric field will again form at its tip, from which a streamer can continue to grow.
Whether streamer continuation occurs therefore depends on the remaining conductivity in the previous channel and on the duration of the second pulse.
Below we derive a rough estimate for streamer continuation.

\begin{figure}
  \centering
  \includegraphics[width=3.5cm]{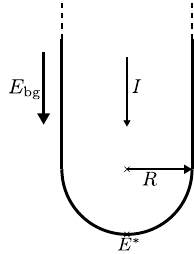}
  \caption{Schematic illustration of the assumed streamer shape used in section~\ref{sec:criterion-streamer-continuation}.
    The streamer is approximated by a cylinder with a hemispherical cap of radius $R$.}
  \label{fig:approximatestreamer}
\end{figure}

We approximate the previous channel at the start of the second pulse as a cylinder with radius $R$ and electric conductivity $\sigma$, with a semi-spherical cap, in which the electric field is equal to the background field $E_\mathrm{bg}$, see figure~\ref{fig:approximatestreamer}.
The electric current in the channel is then initially given by
\begin{equation}
  \label{eq:current}
  I_0 = \pi R^2 \sigma E_\mathrm{bg}.
\end{equation}
This current will decrease over time due to the screening of the background field.
For simplicity, we assume that the accumulated charge at the end of the channel is given by
\begin{equation}
  \label{eq:Q}
  Q(\tau) = I_0 \tau,
\end{equation}
where $\tau$ is the time since the start of the second pulse.
We assume this charge $Q$ gives rise to an electric field similar to that of a (semi-)spherical charge distribution with radius $R$
\begin{equation}
  \label{eq:field}
  E^*(\tau) = E_\mathrm{bg} + \frac{1}{4\pi\varepsilon_0} \frac{Q(\tau)}{R^2}.
\end{equation}
If we now define a required electric field for discharge inception $E_\mathrm{inc}$, we can determine the time $\tau$ it takes until $E^* = E_\mathrm{inc}$.
We include a correction factor $k$ for this time scale, of order unity, which accounts for e.g.\ the partial screening over time and the fact that the actual channel is not a cylinder with a hemispherical cap.
The resulting estimate for $\tau$ is then given by
\begin{equation}
  \label{eq:tau-criterion}
  \tau = 4k \, (E_\mathrm{inc}/E_\mathrm{bg} - 1) \, \tau_\mathrm{drt},
  \quad \mbox{where }\tau_\mathrm{drt} = \varepsilon_0/\sigma.
\end{equation}

Equation \eqref{eq:tau-criterion} provides an estimate for the time $\tau$ needed for a streamer to emerge from the tip of a previous one (as illustrated in figure~\ref{fig:20O2_secondPulseevolution}).
Note that $\tau$ depends on the interpulse time through the conductivity $\sigma$, which decays in the previous channel during the interpulse time.
We will now compare equation \eqref{eq:tau-criterion} against the simulation results presented in section \ref{sec:results}.
We estimate the required field for discharge inception as $E_\mathrm{inc} = 100 \, \textrm{kV/cm}$.
Since $E_\mathrm{bg} = 15 \, \textrm{kV/cm}$, the factor $4k \, (E_\mathrm{inc}/E_\mathrm{bg} - 1)$ is then about $23k$.
In the simulations, we measure the time $\tau_\mathrm{obs}$ until a field of strength $E_\mathrm{inc}$ has been reached at the tip of the previous streamer,
with $\tau_\mathrm{obs} = 0$ corresponding to the start of the second pulse plus its rise time of $1 \, \textrm{ns}$.
Furthermore, we approximate the effective conductivity of the previous streamer as
\begin{equation}
  \label{eq:sigma-def}
  \sigma =  \int_{0}^{R'} 2 \pi r \sigma(r) dr / (\pi R'^2),
\end{equation}
where $\sigma(r)$ is a radial conductivity profile $0.5 \, \textrm{mm}$ behind the streamer head, and $R'$ is defined as the radius at which $\sigma(r)$ has dropped to 10\% of its maximum value.
Since the contribution of ions to the conductivity is relatively small for the cases considered here, we approximate $\sigma(r)$ as $e \mu_e(r) n_e(r)$.

In table~\ref{tab:time-scales}, we list observed continuation times $\tau_\mathrm{obs}$ together with $\tau_\mathrm{drt} = \varepsilon_0/\sigma$, where $\sigma$ was computed according to equation~\eqref{eq:sigma-def} \hm{at the start of the second pulse}.
The ratio $\tau_\mathrm{obs}/\tau_\mathrm{drt}$ lies between 24 and 38, which is in agreement with equation~\eqref{eq:tau-criterion} for $k$ between $1.0$ and $1.5$.

\begin{table*}
    \begin{subtable}{0.5\linewidth}
      \centering
    \begin{tabular}{c|ccc}
      $\tinter$ (ns) &  $\tau_\mathrm{drt}$ (ns) & $\tau_\mathrm{obs}$ (ns) & $\tau_\mathrm{obs}/\tau_\mathrm{drt}$\\
      \hline
      10 & 0.0616 &  1.46 & 23.7 \\
      15 & 0.0978 &  2.86 & 29.2 \\
      20 & 0.162 &  4.88 & 30.1 \\
      25 & 0.267 &  7.01 & 26.3 \\
      50 &- & - & - \\
    \end{tabular}
    \label{tab:continuation-criterion-20o2}
    \caption{20\% O$_2$}
  \end{subtable}
    \begin{subtable}{0.5\linewidth}
      \centering
    \begin{tabular}{c|ccc}
      $\tinter$ (ns) &  $\tau_\mathrm{drt}$ (ns) & $\tau_\mathrm{obs}$ (ns) & $\tau_\mathrm{obs}/\tau_\mathrm{drt}$\\
      \hline
      10 & 0.0349 &  1.11 & 31.8 \\
      15 & 0.0427 &  1.61 & 37.7 \\
      20 & 0.0553 &  2.05 & 37.1 \\
      25 & 0.0699 &  2.52 & 36.1 \\
      50 & 0.164  &  5.77 & 35.2 \\
    \end{tabular}
    \label{tab:continuation-criterion-10o2}
    \caption{10\% O$_2$}
  \end{subtable}
  \caption{Dielectric relaxation times $\tau_\mathrm{drt}$ at the start of the second pulse, based on equation \eqref{eq:sigma-def}, together with the observed inception delays $\tau_\mathrm{obs}$ for second-pulse streamers.
    Specifically, $\tau_\mathrm{obs}$ was here defined as the time it takes for the maximal electric field to reach $100 \, \textrm{kV/cm}$ after the rise time of the second pulse.
  }
  \label{tab:time-scales}
\end{table*}

\subsection{Role of detachment reactions}

The chemistry we use~\cite{baohong_chemistry} has 18 different electron detachment reactions which are of the form
\begin{equation*}
  \text{A}~+~\text{M} \rightarrow~\text{e}~+\text{Products},
\end{equation*}
where A is one of O$^-$, O$_2^-$, O$_3^-$ and M is a neutral species in its ground or excited state.

Table~\ref{tab:ereactions_interpulse} lists the six detachment reactions that contribute significantly to the production of electrons after the end of the first pulse.
Although our chemistry includes electron detachment reactions from O$^-$ ions, they are not included in table \ref{tab:time-scales} because they made a much smaller contribution for our operating conditions.

\begin{figure*}
  \includegraphics[width=16cm]{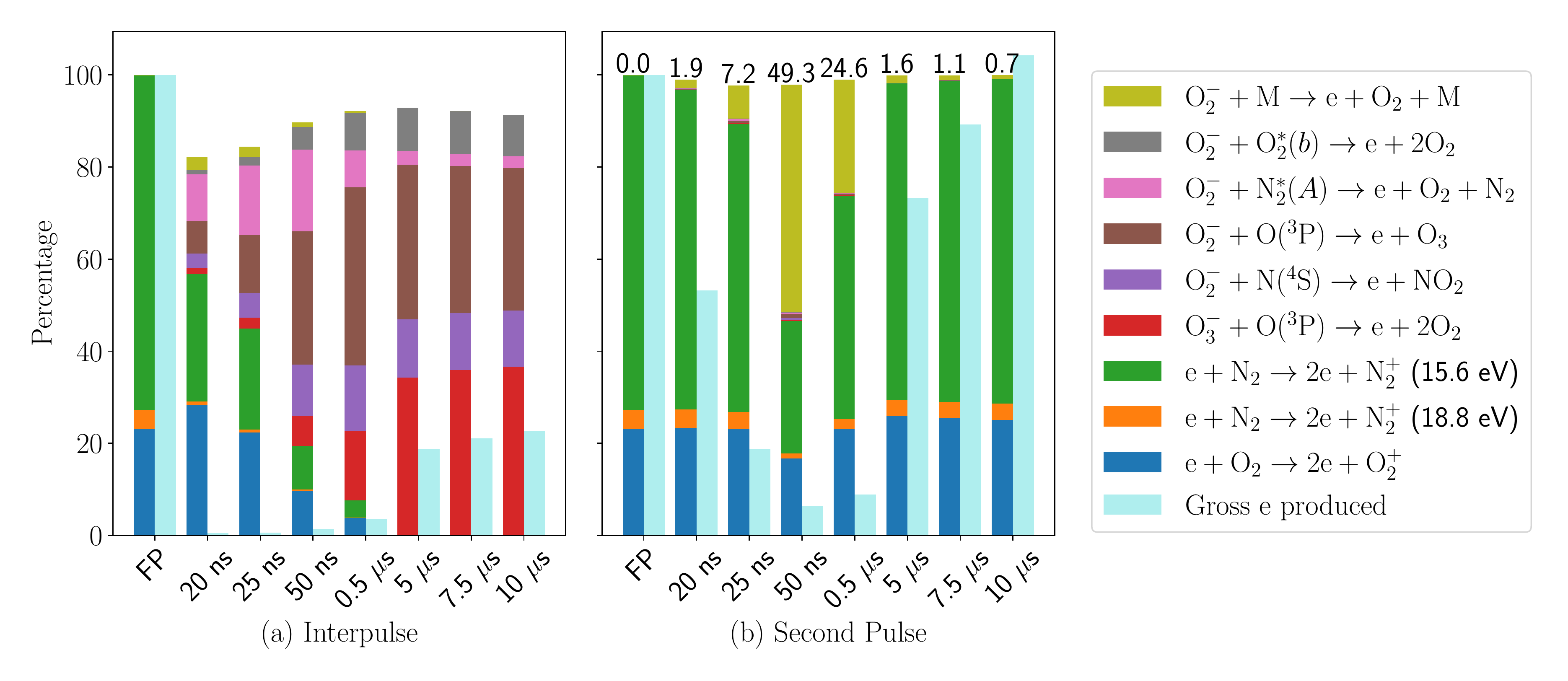}
  \caption{Relative contribution of electron-producing reactions during the interpulse (a) and the second pulse (b) for the case of 20\% O$_2$.
    The production during the first pulse, and the gross electron production normalized with that of the first pulse is shown for comparison.
    The percentage contribution of the field dependent detachment reaction O$_2^-\,+\,\mathrm{M}$ to the gross electron productions detachment is listed at the top of each bar in (b).}
  \label{fig:detachment_20o2}
\end{figure*}

In figure~\ref{fig:detachment_20o2}a we show the relative contribution of electron-producing reactions during the interpulse for the case of 20\% O$_2$.
For short interpulse times, there is still a major contribution from ionization reactions, which take place due to the decaying field enhancement after the voltage is turned off.
For longer interpulse times, most electron production is due to detachment reactions, with the relative contribution of O$_3^-$ detachment increasing over time.
Note that the reaction $\mathrm O_2^- + \mathrm M \to \mathrm e + \mathrm O_2 + \mathrm M$ plays only a minor role during the interpulse due to its field-dependent rate coefficient.



Figure~\ref{fig:detachment_20o2}b shows relative contributions of electron-producing reactions during the second pulse.
Now field-dependent detachment from O$_2^-$ is the major electron detachment reaction. 
For $\tinter$ equal to 50 ns and 500 ns, this reaction is responsible for a significant fraction of the total electron production (which includes ionization reactions).
This is mainly due to two factors: the availability of a large number of O$_2^-$ ions, and a reduction in ionization reactions due to lower field enhancement in the inhibited growth regime, see figure~\ref{fig:20O2_secondPulseevolution}b.
For interpulse times of $5 \, \mu\textrm{s}$ or more, O$_2^-$ ions are lost due to negative-ion conversion and ion-ion recombination reactions.
For interpulse times below $25 \, \textrm{ns}$, detachment from O$_2^-$ does not play a significant role since the previous channel is quickly screened and a new streamer forms.\\
Results for the case of 10\% O$_2$ and 90\% N$_2$ are given in appendix~\ref{sec:detachment-comparision-varyo2}.
We observe a similar trend in the relative contributions of detachment reactions for this case.

\subsection{Effect of using a different chemistry set}
\label{sec:diffchemset}

For comparison, we have also simulated the electron density decay during the interpulse with a simpler plasma chemistry from~\cite{li2022computational}.
This chemistry contains 8 species and 15 reactions, and it lacks the extensive electron, ion, neutral and excited species reactions as compared to the chemistry set used in the rest of this paper.
For electrons, the same Phelps~\cite{Phelps_database} cross sections are used.
In figure~\ref{fig:interpulse_pvt}, where we denote the smaller chemistry from this section as `Simple Chemistry', we compare the electron decay inside the streamer channel during the interpulse phase.
The decay agrees well up to 100 ns, but at later times the simpler chemistry predicts a much faster decay due to the absence of the various electron detachment reactions.
For a pulse duration of 100 ns or more, the sets would therefore lead to different streamer continuation times.





\section{Conclusions}

We have investigated streamer continuation with double-pulse simulations in \hm{N$_2$} containing 20\% and 10\% \hm{O$_2$} at 1~bar.
We use two identical voltage pulses of 15\,ns, and vary the interpulse time $\tinter$ between 5\,ns and 10 $\mu$s.
For increasing $\tinter$, we observe three regimes during the second pulse:
\begin{itemize}
  \item In the streamer continuation regime ($\tinter \lesssim 50\,\mathrm{ns}$), a new streamer emerges from the tip of the previous one.
  \item In the inhibited growth regime ($50 \, \mathrm{ns} \lesssim \tinter \lesssim 500\,\mathrm{ns}$, with somewhat longer time scales for 10\% O$_2$), the previous channel is partially re-ionized, but there is considerable less field enhancement and almost no light emission.
  This re-ionization wave becomes stronger for increasing interpulse times, characterized by increased electric field screening and increased light emission.
  \item For $\tinter \gtrsim 5\,\mu\mathrm{s}$, a new streamer forms that is similar to the first one.
  However, even for the longest interpulse times considered here ($10 \, \mu\textrm{s}$) this new streamer has lower field enhancement and is less bright than the first one.
\end{itemize}
With 10\% O$_2$ these regimes occured at slighly longer interpulse times than with 20\% O$_2$, due to lower electron attachment and recombination rates.

In our simulations, streamer continuation occurred when the remaining electron density was in the range of $10^{19}  \, \mathrm{m}^{-3}$ up to $4\times 10^{19} \, \mathrm{m}^{-3}$.
This range agrees with the estimate made in~\cite{nijdam_investigation_2014}, based on experimental measurements and 0D modeling, when scaled to the same pressure.
We derived an estimate for the time needed for streamer continuation to occur which depends on the conductivity of the previous channel and the background electric field.
This estimate and our observed values of streamer inception delay in the simulations are in reasonable agreement.
Furthermore, we show that for interpulse times above 100 ns several electron detachment reactions significantly slow down the decay of the electron density.






\section{Outlook}

We obtained our current results using a specific background electric field, pulse shape and pressure.
In future work, our computational model could be improved in the following aspects:
\begin{itemize}
  \item When the voltage falls to zero, the polarity of the electrode is reversed due to the remaining positive space charge. A cathode sheath with a really high local field is expected to form around the electrode, see e.g.~\cite{Odrobina_1995,Babaeva_2015b,Yan_2014a}.
  Resolving such a sheath was computationally not feasible with our model, which is why homogeneous Neumann boundary conditions for species at the electrodes were used.
  The effect of more realistic electrode boundary conditions could be studied in future work.
  \item As shown in section~\ref{sec:diffchemset}, for longer interpulse times, it is important that we use a chemistry set that is complete and has accurate rate constants.
  It was also pointed out in~\cite{baohong_chemistry} that some rate constants also depend on pressure and temperature.
  Thus, in order to be able to extend our results to different pressures, future work involves generalizing the reaction set.
  \item Finally, varying the pulse shape, background electric field, and gas composition might result in streamer branching, which would require fully 3D modeling.
\end{itemize}
These present and future investigations will allow for theoretical predictions for optimal pulse repetition frequencies for various plasma processing applications.

\section*{Acknowledgements}

This work was carried out on the Dutch national e-infrastructure with the support of SURF Co-operative. H.M. was supported by the STW-project 15052 ``Let $\text{CO}_2$ spark". A.M. has received funding from the European Union’s Horizon 2020 research and innovation program under project 722337 (SAINT). We thank Baohong Guo for his help with the plasma chemistry used in this paper.

\section*{Data Availability}
The data that support the findings of this study are openly available in the following URL/DOI: \url{https://doi.org/10.5281/zenodo.8055893}.
\appendix

\section{Effect of varying O$_2$ concentration on detachment reactions}
\label{sec:detachment-comparision-varyo2}
\begin{figure*}
  \includegraphics[width=16cm]{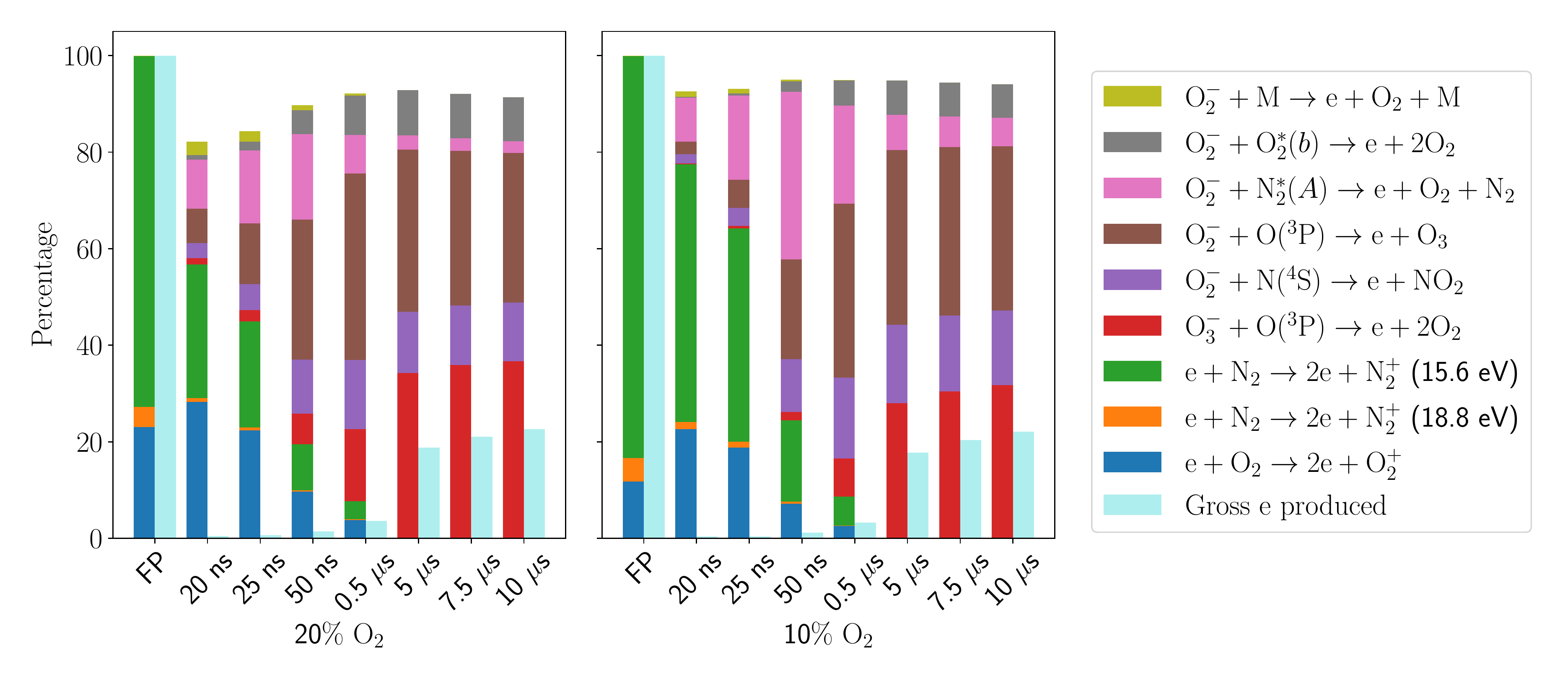}
  \caption{Relative contribution of electron-producing reactions during the interpulse for 10\% O$_2$.
    The production during the first pulse is shown for comparison. 
    We also show the gross electron production normalized to that of the first pulse.}
  \label{fig:detachment_interpulse_10O2}
\end{figure*}

\begin{figure*}
  \includegraphics[width=16cm]{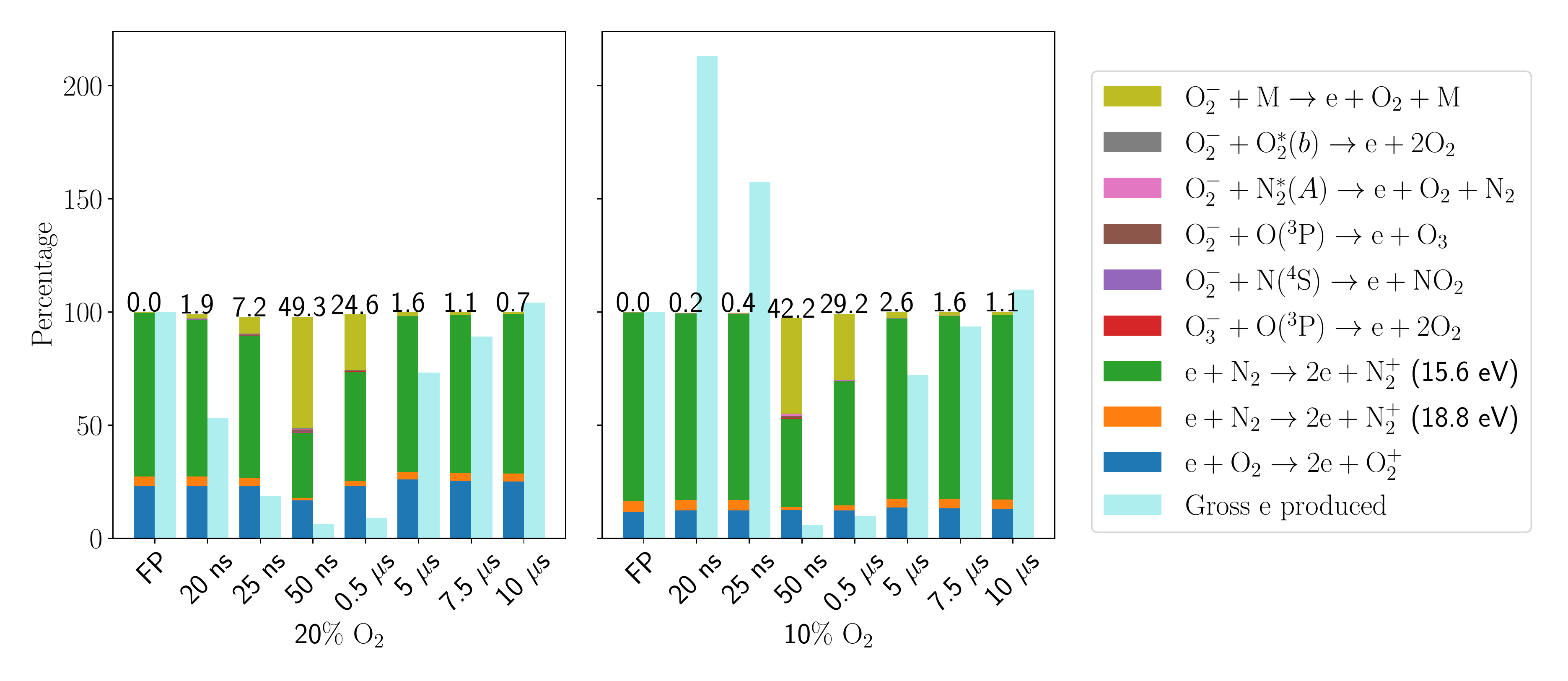}
  \caption{Relative contribution of electron-producing reactions during the second pulse for 10\% O$_2$.
    The production during the first pulse and the gross electron production normalized to that of the first pulse is also shown.
    The percentage contribution of the field-dependent detachment reaction O$_2^-\,+\,\mathrm{M}$ to the gross electron productions detachment is listed at the top of each bar plot.
    For 10\% O$_2$ and $\tinter = 20\,\mathrm{ns}$ and $\tinter = 25\,\mathrm{ns}$, gross electron production during the second pulse is higher than during the first pulse.
  }
  \label{fig:detachment_secondPulse_10O2}
\end{figure*}

Here we show the relative contribution (in \%) of various electron-producting reactions for the case of 10\% O$_2$. 
Figure~\ref{fig:detachment_interpulse_10O2} shows the contribution during the interpulse phase and figure~\ref{fig:detachment_secondPulse_10O2} shows the contribution during the second pulse.
In each of these figures, we also show the results from figure~\ref{fig:detachment_20o2} for comparison.

\section{Electric field inside the streamer channel during the interpulse}
\label{sec:electric_field_interpulse}

\begin{figure}
  \centering
  \includegraphics[width=8cm]{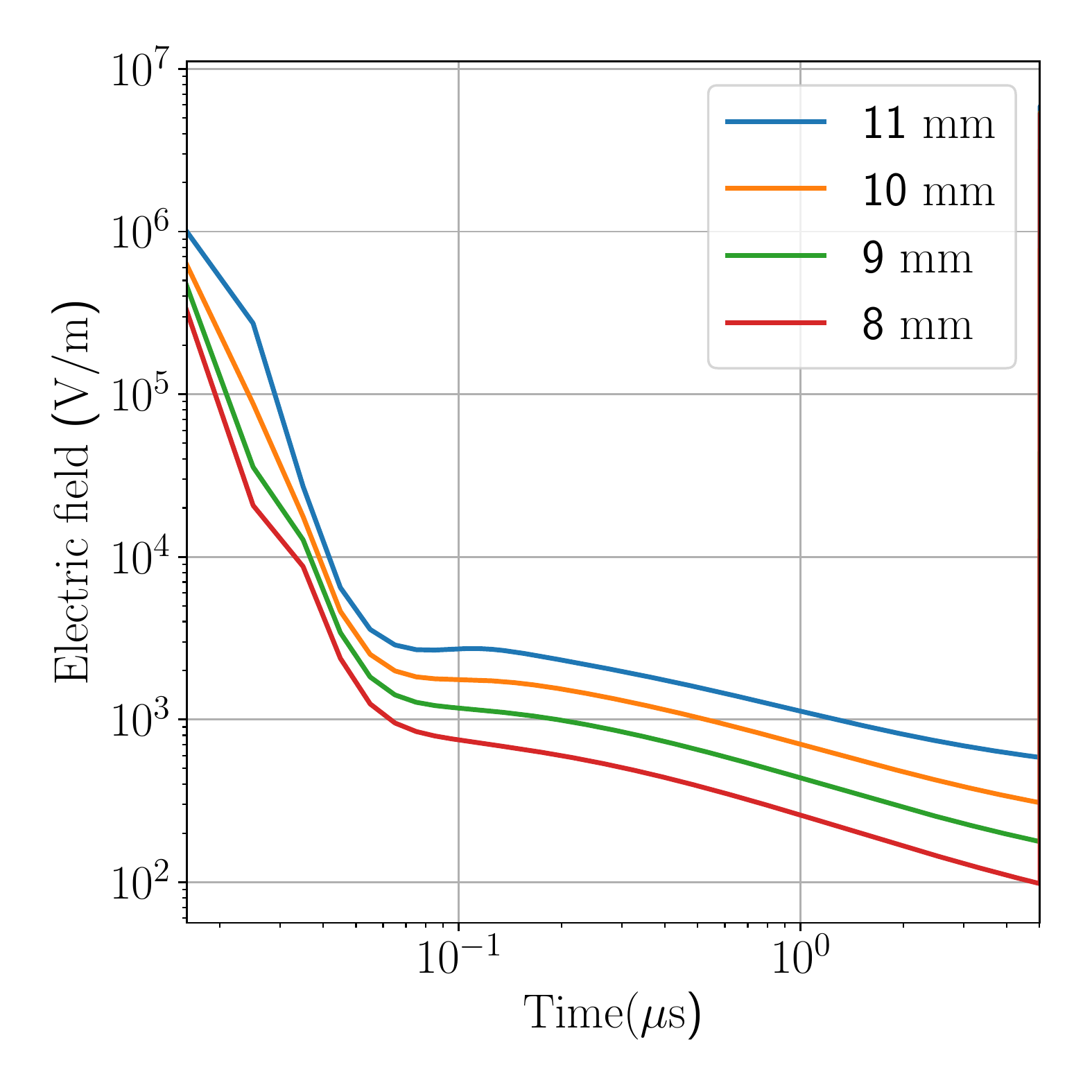}
  \caption{The temporal evolution of the on-axis electric field inside the streamer channel during the interpulse phase.
    The values were obtained at points $1$, $2$, $3$ and $4 \, \textrm{mm}$ below the electrode tip, which is approximately located at $z = 12 \, \mathrm{mm}$.
  }
  \label{fig:interpulse_efld}
\end{figure}

\hm{In Figure~\ref{fig:interpulse_efld}, we show the electric field decay over time at various locations along the axis of the streamer channel for 20\% O$_2$.
The electric field decays from $10^6$\,V/m (41 Td) to $10^2$\,V/m (0.004 Td) over a time interval of 5\,$\mu$s.
However, the boundary conditions at the electrode are currently not very realistic in our model: a Neumann zero condition is used for electrons, which results in electrons freely entering the domain after the voltage is turned off.
This leads to a faster decay of the residual charge (and thus also of the electric field) than would occur with more realistic boundary conditions.

In principle it is possible to specify more realistic boundary conditions at the electrode, in which the outflow of electrons is caused by secondary emission processes.
However, with such boundary conditions we cannot run simulations for long time scales, since the anode effectively becomes a cathode after the voltage is turned off, resulting in the formation of thin sheaths with very high local electric fields.
We intend to improve our model's boundary conditions in the future, for example by using the local energy approximation as in~\cite{jovanovic2022streamer} and/or by approximating the electron dynamics in the sheath.}

\section{Effect of gas heating}
\label{sec:gas_heating}
\begin{figure}
        \centering
        \includegraphics[width=8cm]{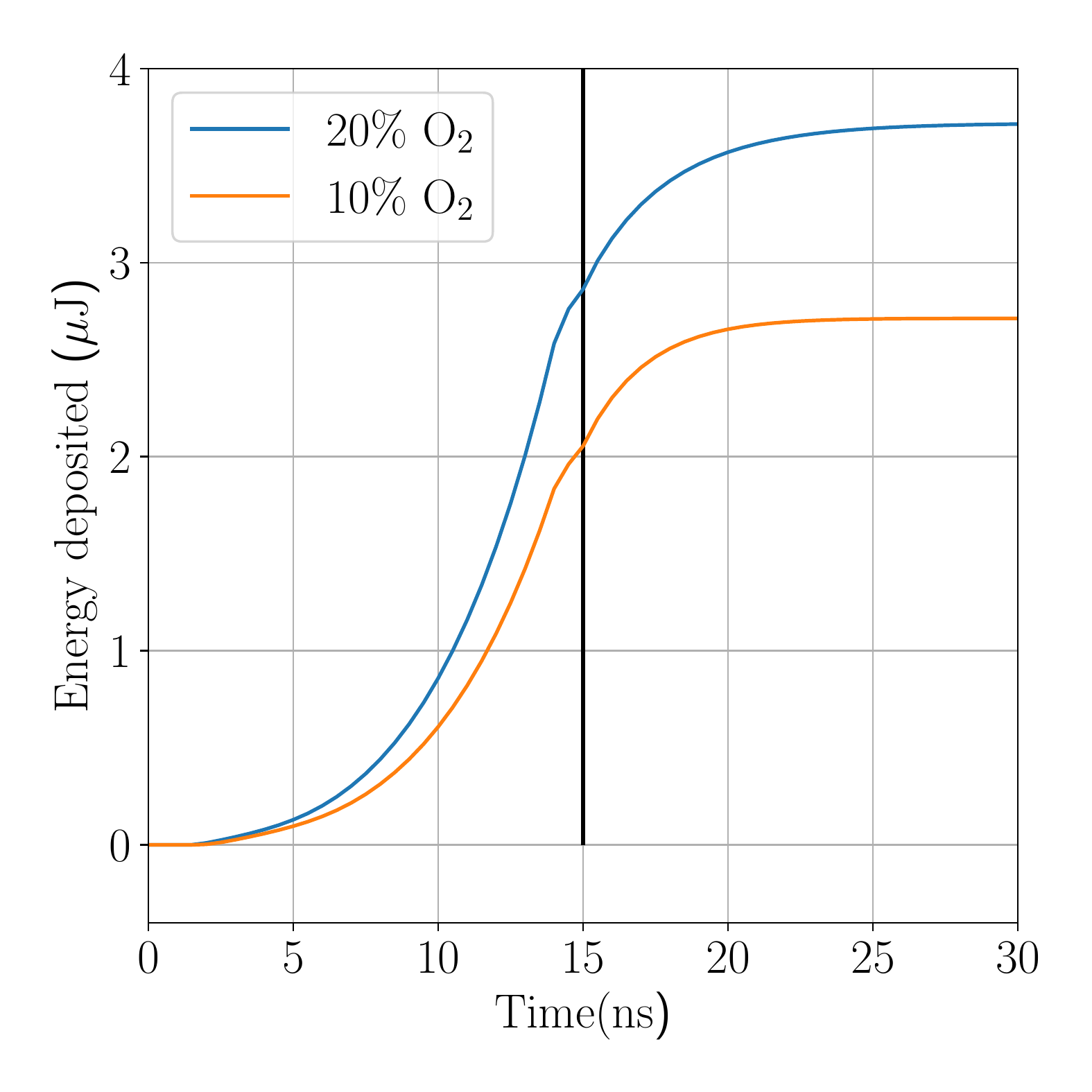}
        \caption{Space-time integrated Joule heating term. The vertical line is at the 15\,ns mark, where the first pulse is turned off.}
        \label{fig:energydep}
    \end{figure}

\hm{Figure~\ref{fig:energydep} shows the space-time integrated Joule heating ($\vec{j} \cdot \vec{E}$) during the first pulse.
  The energy deposited due to the first pulse is below 4\,$\mu \mathrm{J}$, and no energy is deposited after 25 ns.
  At 1 bar and 300 K, dry air (80\% N$_2$:20\% O$_2$) has a density $\rho=1.2\,\mathrm{kg} \, \mathrm{m}^{-3}$ and a specific heat capacity $C_p= 1000 \, \mathrm{J}\,\mathrm{kg}^{-1}\,\mathrm{K}^{-1}$.
  This energy is deposited in a cylindrical volume of radius $r=0.8$ mm (maximum width of the streamer channel) and height $h \approx 5.5$ mm (first pulse streamer length).
  Even if we assume that all the energy instantaneously goes into gas heating, the temperature increase is only about $\Delta \mathrm{T} = 4 \mu \mathrm{J}/(C_p \rho \pi r^2 h) \approx 0.3 \, \mathrm{K}$.
  Such a small increase in temperature will not have a significant effect on the second-pulse streamer properties, which is why gas heating was not included in our model.
  In a follow-up paper that is in preparation, we are exploring the effects of gas heating after many pulses.}

\section*{References}
\bibliography{References}

\providecommand{\newblock}{}
\begin{thebibliography}{10}
\expandafter\ifx\csname url\endcsname\relax
  \def\url#1{{\tt #1}}\fi
\expandafter\ifx\csname urlprefix\endcsname\relax\def\urlprefix{URL }\fi
\providecommand{\eprint}[2][]{\url{#2}}

\bibitem{vitello1994simulation}
Vitello P, Penetrante B and Bardsley J 1994 {\em Physical Review E\/} {\bf 49}
  5574

\bibitem{nijdam_physics_2020}
Nijdam S, Teunissen J and Ebert U 2020 {\em Plasma Sources Sci. Technol.\/}
  {\bf 29} 103001 ISSN 0963-0252 publisher: IOP Publishing
  \urlprefix\url{https://doi.org/10.1088/1361-6595/abaa05}

\bibitem{verloy_cold_2020}
Verloy R, Privat-Maldonado A, Smits E and Bogaerts A 2020 {\em Cancers\/} {\bf
  12} 2782 ISSN 2072-6694
  \urlprefix\url{https://www.mdpi.com/2072-6694/12/10/2782}

\bibitem{graves_low_2014}
Graves D~B 2014 {\em Physics of Plasmas\/} {\bf 21} 080901 ISSN 1070-664X
  publisher: American Institute of Physics
  \urlprefix\url{https://aip.scitation.org/doi/abs/10.1063/1.4892534}

\bibitem{laroussi_2014_medicine}
Laroussi M 2014 {\em Plasma Processes and Polymers\/} {\bf 11} 1138--1141
  \urlprefix\url{https://onlinelibrary.wiley.com/doi/abs/10.1002/ppap.201400152}

\bibitem{ranieri_plasma_2021}
Ranieri P, Sponsel N, Kizer J, Rojas-Pierce M, Hernández R, Gatiboni L,
  Grunden A and Stapelmann K 2021 {\em Plasma Processes and Polymers\/} {\bf
  18} 2000162
  \urlprefix\url{https://onlinelibrary.wiley.com/doi/abs/10.1002/ppap.202000162}

\bibitem{bardos2010cold}
B{\'a}rdos L and Bar{\'a}nkov{\'a} H 2010 {\em Thin solid films\/} {\bf 518}
  6705--6713

\bibitem{starikovskaia2006plasma}
Starikovskaia S~M 2006 {\em Journal of Physics D: Applied Physics\/} {\bf 39}
  R265

\bibitem{nijdam_investigation_2014}
Nijdam S, Takahashi E, Markosyan A~H and Ebert U 2014 {\em Plasma Sources Sci.
  Technol.\/} {\bf 23} 025008

\bibitem{li2018positive}
Li Y, Van~Veldhuizen E~M, Zhang G~J, Ebert U and Nijdam S 2018 {\em Plasma
  Sources Science and Technology\/} {\bf 27} 125003

\bibitem{mirpour_distribution_2020}
Mirpour S, Martinez A, Teunissen J, Ebert U and Nijdam S 2020 {\em Plasma
  Sources Science and Technology\/} {\bf 29} 115010
  \urlprefix\url{https://doi.org/10.1088/1361-6595/abb614}

\bibitem{pai2010transitions}
Pai D~Z, Lacoste D~A and Laux C~O 2010 {\em Journal of Applied Physics\/} {\bf
  107} 093303

\bibitem{tholin2013simulation}
Tholin F and Bourdon A 2013 {\em Plasma Sources Science and Technology\/} {\bf
  22} 045014

\bibitem{lu2018guided}
Lu X and Ostrikov K 2018 {\em Applied Physics Reviews\/} {\bf 5} 031102

\bibitem{chang2016effect}
Chang L, Nie L, Xian Y and Lu X 2016 {\em Physics of Plasmas\/} {\bf 23} 123513

\bibitem{babaeva2022evolution}
Babaeva N~Y, Naidis G, Tarasenko V, Sorokin D, Zhang C and Shao T 2022 {\em
  Plasma Science and Technology\/}

\bibitem{hoft2014bidirectional}
H{\"o}ft H, Kettlitz M, Weltmann K and Brandenburg R 2014 {\em Journal of
  Physics D: Applied Physics\/} {\bf 47} 455202

\bibitem{hoft2014breakdown}
H{\"o}ft H, Kettlitz M, Becker M, Hoder T, Loffhagen D, Brandenburg R and
  Weltmann K 2014 {\em Journal of Physics D: Applied Physics\/} {\bf 47} 465206

\bibitem{Zhao_2020}
Zhao Z, Huang D~D, Wang Y~N, Li C~J and Li J~T 2020 {\em Plasma Sources Science
  and Technology\/} {\bf 29} 015016 ISSN 1361-6595

\bibitem{Zhao_2023}
Zhao Z, Li C, Guo Y, Zheng X, Sun A and Li J 2023 {\em Plasma Sources Science
  and Technology\/} {\bf 32} 015002 ISSN 0963-0252, 1361-6595

\bibitem{Zhao_2020a}
Zhao Z and Li J 2020 {\em High Voltage\/} {\bf 5} 569--582 ISSN 2397-7264,
  2397-7264

\bibitem{nijdam_streamer_2014}
Nijdam S, Takahashi E, Teunissen J and Ebert U 2014 {\em New J. Phys.\/} {\bf
  16} 103038 ISSN 1367-2630

\bibitem{Nijdam_2016a}
Nijdam S, Teunissen J, Takahashi E and Ebert U 2016 {\em Plasma Sources Science
  and Technology\/} {\bf 25} 044001 ISSN 0963-0252, 1361-6595

\bibitem{pancheshnyi2005role}
Pancheshnyi S 2005 {\em Plasma Sources Science and Technology\/} {\bf 14} 645

\bibitem{naidis2011modelling}
Naidis G 2011 {\em Journal of Physics D: Applied Physics\/} {\bf 44} 215203

\bibitem{teunissen_simulating_2017}
Teunissen J and Ebert U 2017 {\em Journal of Physics D: Applied Physics\/} {\bf
  50} 474001 \urlprefix\url{https://doi.org/10.1088/1361-6463/aa8faf}

\bibitem{li_comparing_2021}
Li X, Dijcks S, Nijdam S, Sun A, Ebert U and Teunissen J 2021 {\em Plasma
  Sources Science and Technology\/} {\bf 30} 095002
  \urlprefix\url{https://doi.org/10.1088/1361-6595/ac1b36}

\bibitem{Wang_2022}
Wang Z, Sun A and Teunissen J 2022 {\em Plasma Sources Science and
  Technology\/} {\bf 31} 015012 ISSN 0963-0252, 1361-6595

\bibitem{Bagheri_2018a}
Bagheri B, Teunissen J, Ebert U, Becker M~M, Chen S, Ducasse O, Eichwald O,
  Loffhagen D, Luque A, Mihailova D, Plewa J~M, {van Dijk} J and Yousfi M 2018
  {\em Plasma Sources Science and Technology\/} {\bf 27} 095002 ISSN 1361-6595

\bibitem{hagelaar_solving_2005}
Hagelaar G~J~M and Pitchford L~C 2005  {\bf 14} 722--733 ISSN 0963-0252
  publisher: IOP Publishing
  \urlprefix\url{https://doi.org/10.1088/0963-0252/14/4/011}

\bibitem{Phelps_database}
{Phelps} database, www.lxcat.net, retrieved on august 19, 2021.

\bibitem{zheleznyak1982photoi_english}
{Zhelezniak} M~B, {Mnatsakanian} A~K and {Sizykh} S~V 1982 {\em High
  Temperature Science\/} {\bf 20} 357--362

\bibitem{Teunissen_2018_afivo}
Teunissen J and Ebert U 2018 {\em Computer Physics Communications\/} {\bf 233}
  156--166 ISSN 0010-4655

\bibitem{Teunissen_2023}
Teunissen J and Schiavello F 2023 {\em Computer Physics Communications\/} {\bf
  286} 108665 ISSN 00104655

\bibitem{baohong_chemistry}
Guo B and Teunissen J 2023 {\em Plasma Sources Science and Technology\/} {\bf
  32} 025001 ISSN 0963-0252, 1361-6595

\bibitem{kossyi_kinetic_1992}
Kossyi I~A, Kostinsky A~Y, Matveyev A~A and Silakov V~P 1992 {\em Plasma
  Sources Sci. Technol.\/} {\bf 1} 207--220 ISSN 0963-0252, 1361-6595
  \urlprefix\url{https://iopscience.iop.org/article/10.1088/0963-0252/1/3/011}

\bibitem{ono2020}
Ono R and Komuro A 2020 {\em Journal of Physics D: Applied Physics\/} {\bf 53}
  035202 ISSN 0022-3727, 1361-6463

\bibitem{tochikubo2002numerical}
Tochikubo F and Arai H 2002 {\em Japanese journal of applied physics\/} {\bf
  41} 844

\bibitem{li2022computational}
Li X, Guo B, Sun A, Ebert U and Teunissen J 2022 {\em Plasma Sources Science
  and Technology\/} {\bf 31} 065011

\bibitem{pancheshnyi_development_2005}
Pancheshnyi S, Nudnova M and Starikovskii A 2005 {\em Phys. Rev. E\/} {\bf 71}
  016407 publisher: American Physical Society
  \urlprefix\url{https://link.aps.org/doi/10.1103/PhysRevE.71.016407}

\bibitem{hansen1985recursive}
Hansen E~W and Law P~L 1985 {\em JOSA A\/} {\bf 2} 510--520

\bibitem{aleksandrov_ionization_1999}
Aleksandrov N~L and Bazelyan E~M 1999 {\em Plasma Sources Science and
  Technology\/} {\bf 8} 285--294
  \urlprefix\url{https://doi.org/10.1088/0963-0252/8/2/309}

\bibitem{pancheshnyi2013effective}
Pancheshnyi S 2013 {\em Journal of Physics D: Applied Physics\/} {\bf 46}
  155201

\bibitem{Bolsig_application}
 2019 Bolsig+ (linux) \urlprefix\url{http://www.bolsig.laplace.univ-tlse.fr/}

\bibitem{Odrobina_1995}
Odrobina I and {\v C}ern{\'a}k M 1995 {\em Journal of Applied Physics\/} {\bf
  78} 3635--3642 ISSN 0021-8979, 1089-7550

\bibitem{Babaeva_2015b}
Babaeva N~Y 2015 {\em Plasma Sources Science and Technology\/} {\bf 24} 034012
  ISSN 0963-0252, 1361-6595

\bibitem{Yan_2014a}
Yan W, Liu F, Sang C and Wang D 2014 {\em Physics of Plasmas\/} {\bf 21} 013504
  ISSN 1070-664X, 1089-7674

\bibitem{jovanovic2022streamer}
Jovanovi{\'c} A~P, Loffhagen D and Becker M~M 2022 {\em Plasma Sources Science
  and Technology\/} {\bf 31} 04LT02

\end{thebibliography}
\end{document}